\documentclass[twocolumn,showpacs,preprintnumbers,amsmath,amssymb]{revtex4}

\usepackage{dcolumn}
\usepackage{bm}
\usepackage{epsfig,graphicx}
\usepackage{times}

\begin{document}

\title{Walks on Apollonian networks}

\author{Zi-Gang Huang$^{1}$, Xin-Jian Xu$^{2}$, Zhi-Xi Wu$^{1}$, and
Ying-Hai Wang$^{1,}$\footnote{For correspondence: yhwang@lzu.edu.cn}}
\affiliation{$^{1}$Institute of Theoretical Physics, Lanzhou
University, Lanzhou Gansu 730000, China\\
$^{2}$Department of Electronic Engineering, City University of
Hong Kong, Kowloon, Hong Kong, China}

\date{\today}

\begin{abstract}
We carry out comparative studies of random walks on deterministic Apollonian networks (DANs) and random Apollonian networks (RANs). We perform computer simulations for the mean first passage time, the average return time, the mean-square displacement, and the network coverage for unrestricted random walk. The diffusions both on DANs and RANs are proved to be sublinear. The search efficiency for walks with various strategies and the influence of the topology of underlying networks on the dynamics of walks are discussed. Contrary to one's intuition, it is shown that the self-avoiding random walk, which has been verified as an optimal strategy for searching on scale-free and small-world networks, is not the best strategy for the DAN in the thermodynamic limit.
\end{abstract}

\maketitle

In the past few years, much attention has been devoted
to the characterization and modelling of a wide range of complex
systems that can be described as networks
\cite{AlbertDease,Dorogovtsev,Newman}. The topological properties
of real-world networks have been studied extensively. But an even
more intriguing task, and a natural extension of these studies, is
to understand how the topological structure of networks affects
dynamics taking place on top of them \cite{Strogatz}. Many
dynamic processes have been studied on complex networks, such as
epidemic spreading \cite {May}, percolation \cite {Callaway}, and
synchronization \cite {Barahona} \emph{et al}. These researches
show that topologies of networks play a crucial role in determining
the system dynamical features.

Random walk has been used for modelling various dynamics in
physical, biological, and social contexts \cite{Spitzer}. It could
also be a mechanism of transport and search on network
\cite{Adamic,Guimer,Holme}. Thus, one interesting problem is to
study the dynamical behavior of a random walker on networks with
different topological properties. Much is known about random walks
on both regular and random networks \cite{Barber,Hughes}.
Recently, there have also been several studies of random walks on
small-world networks (SWNs) \cite{Pandit,Lahtinen,Almaas,Parris}
and on scale-free networks (SFNs) \cite{Adamic,Gallos,Noh,Yang}.
The impacts of the heterogeneous topological structures of the
networks on the nature of the diffusive and relaxation dynamics of
the random walk were probed \cite{Adamic,Jespersen,Tadic,Guimer}.
Also, finding efficient algorithms for searching on networks with
different topological properties is an important issue related to
random walks \cite{Adamic,Noh,Yang}.

In this paper we investigate walking processes taking place upon
the determinist Apollonian network (DAN) \cite{Andrade} and its variation, the
random Apollonian network (RAN) \cite{Zhou}. The DAN can be defined
based of the ancient problem of filling space with spheres, first
tackled by the Greek mathematician Apollonius of Perga \cite{Boyd}.
That is, starting with an initial array of touching disks, which have curvilinear-triangle interstices, disks are added inside each existed interstice in the present configuration, such that these disks touch each of the disks bounding the curvilinear triangles. These added disks give rise to three smaller interstices, which will be filled in the next generation. This process is then repeated for successive generations. The DAN is constructed based on this process by considering each disk as a node, and the disks in contact as the corresponding nodes connected. For each new node added to a certain triangle (corresponding to the curvilinear-triangle interstice), the three vertices are linked to, and three new triangles are created in the network, into which nodes will be inserted in the next generation. Different from recursive constructing of the DAN \cite{Andrade}, the RAN starts with a triangle containing three nodes. Then, at each time step, one triangle is randomly selected to add a new node \cite{Zhou}. Both networks are simultaneously scale-free, small-world, Euclidean, and space filling, and have attracted an increasing interest recently \cite{Doye,Andrade_1,Zhang}.

We carry out the random walk along the bonds of a given network as
follows: (i) There is only one walker on the network at a time;
(ii) The random walker is injected onto a randomly chosen node on
the network, a new node for each walker. We will call this node
the ``origin'' of the walk; (iii) At each discrete time step $t$,
the walker will jump to one nearest neighbor of its current node
according to one certain strategy; (iv) We average over different
random walkers and realizations of the network until the results
converge.

The walk strategies adopted by the walker include the following:
random walk (RW), no-back (NB) walk, no-triangle-loop (NTL) walk,
no-quadrangle-loop (NQL) walk, and self-avoiding (SA) random walk.
For the RW, the walker may unrestrictedly jump to a neighbor node
by randomly taking one of the links. It forgets all information
about its past. The NB walk implies that a random walker, if
possible, will not return to the node it was situated at the
previous step. Similarly, NTL and NQL random walks mean that the
walker will try to avoid walking in loops, with three or four
edges, respectively, unless there is no other choice. We mention
that the NQL walk also includes the NTL, which means it eliminates
quadrangle loops as well as triangle loops. Finally, the SA random
walk implies that the walker is the smarter. It tries to avoid
revisiting the node that it has already visited in a run of the
search. Surely, the SA walk also includes the NTL and NQL walks.

One can imagine that an unrestricted RW walker may be trapped into
high-clustering regions and revisit nodes there for a long time by reason of
walking in loops. While a walker adopting other strategies like NB, NTL and NQL
can escape from those regions easier, and therefore improve the search
efficiency. Thus, different clustering properties of the underlying networks
will induce different efficiencies of the above mentioned strategies. In return,
the clustering effect of the networks could be reflected by the efficiencies of
those search strategies.

\begin{figure}
\includegraphics[width=8.4cm]{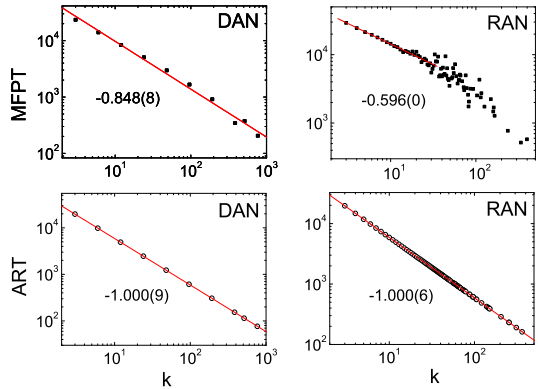}

\caption{(Color online) The mean first passage times $\langle{T_{ij}}\rangle$ (upper panels) and the average return time $\langle{T_{ii}}\rangle$ (lower panels), both of which are average over the results of given $k$'s, for DANs (left-hand) and RANs (right-hand). Fitted linear relations of average return time (solid lines) are obtained with a slope of $-1$ for both DANs and RANs.} \label{fig1}
\end{figure}

In the context of transport and search, the mean first passage
time (MFPT) and the average return time (ART) are important
characteristics of the random walk. The average return time
denoted by $\langle{T_{ii}}\rangle$ is the average time needed for
a walker to return to the origin $i$, which is a
special case of MFPT from $i$ to $j$ denoted by $\langle{T_{ij}}\rangle$.
Noh and Rieger \cite{Noh} have proved that the MFPT is negatively
correlated with $K_{j}$, and $\langle{T_{ii}}\rangle$ is determined only
by the total number of links and the degree of the node $i$
\begin{equation}
\langle{T_{ii}}\rangle=\mathcal{N}/K_{i}\label{eq1}
\end{equation}
with $\mathcal{N}=\sum_{j}{K_{j}}$ $(j=1,\ldots,N)$. Namely, nodes with higher
degrees are visited earlier and more frequently, and therefore targets on these
nodes are more easily found out than on nodes with smaller degrees. In Fig.
\ref{fig1} we depict simulations of the MFPT (the upper panels) and ART
(the lower panels) as a function
of nodes degree for the DAN and RAN with size $N=9844$. After averaging over
the MFPT and ART for a given degree $k$, the linear property of the log-log
plot of ART (the lower panels) is in excellent agreement with Eq. (\ref{eq1}), and
the negative correlation between MFPT and $k$ are also presented (the upper panels).

In the following, we will investigate walking processes on DANs and RANs. The
corresponding results on  Watts-Strogatz (WS) ($K=3$, $p_{0}=0.1$) \cite{Watts} and Barab\'{a}si-Albert (BA) ($m=m_{0}=3$) \cite{Barabasi} networks are also presented
for comparison. Note that, the four networks have the same average degree $\langle{k}\rangle=6$.

\begin{figure*}
\includegraphics[width=15cm]{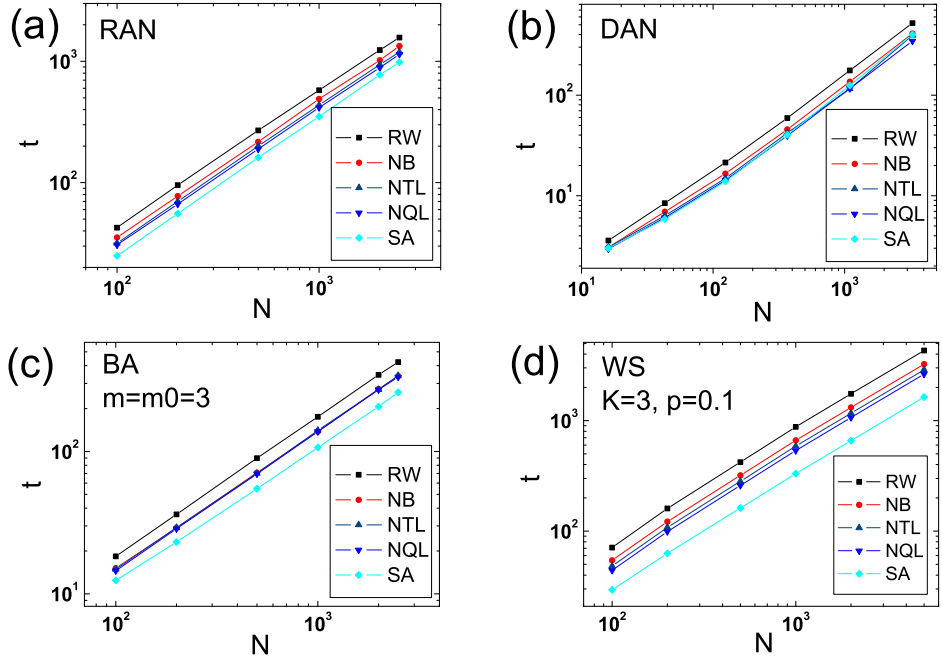}
\caption{(Color online) Average search times of various search
strategies for RANs (a), DANs (b), BA networks (c), and WS
networks (d).} \label{fig2}
\end{figure*}

Firstly, let us discuss the search efficiency of all the five walk strategies mentioned
above for the four networks. We suppose at every step, the walker adopting various strategies  only know neighbors of its present node. So if the target is at one neighbor of the present node where the walker stays, this round of search is over. The walks for the five strategies are performed and the search times (see Fig. \ref{fig2}) are averaged over $200$ different couples of initial positions and targets of the walker, and $50$ realizations of the networks. One can find some common properties from Fig. \ref{fig2}. For example, the approximate
power-law relations hold for all search processes of RW, NB, NTL, NQL and SA
walks. As another example, the RW walk is the most inefficient search strategy
for all the four networks. Nevertheless, some obvious differences still exist
among them. The search efficiencies of NB, NTL and NQL walks nearly collapse
into one only for BA network (see Fig. \ref{fig2} (c), also in Ref. \cite{Yang}).
This can be understood by its comparatively smallest clustering coefficient as
to the other three networks. In the weak clustering networks, the NB, NTL and
NQL walkers, who adopt increasingly strict law to avoid revisit, can not take
their advantages and their search efficiencies keep very closely or even
collapse into one. Hence, for DANs and RANs, the efficiency-improvement of NTL
and NQL walks as to the NB walk reflects the fact that high clustered nodes are
popular on them.

\begin{table}
\caption{\label{tab:table1} The PSP threshold $P_{c}$ and average path length $\langle{L}\rangle$ of the four networks with size $N=9844$.}
\begin{ruledtabular}
\begin{tabular}{cdddd}
&           \mbox{DAN}&    \mbox{RAN}&      \mbox{BA}&    \mbox{WS}\\
\hline
$P_{c}\approx$      & 4/N       & 0.03      & 0.3       & 0.63      \\
$\langle{L}\rangle$ &4.06(5)    &5.40(9)    & 4.27(9)   & 8.61(9)   \\
\end{tabular}
\end{ruledtabular}
\end{table}

The SA walk was proved to be generally the most efficient strategy for
searching on networks, whose small-world properties were considered to be the
reason \cite{Yang}. However, it is counterexample for the DAN (see Fig.
\ref{fig2}(b)). The DAN is one of typical regular networks which has
deterministic size when the number of the generations is certain
\cite{Andrade,Doye}. Its short average length and high clustering coefficient
indeed describe a small-world scenario. Fig. \ref{fig2}(b) shows the average
search times for DANs with $3$ to $8$ generations, corresponding to network
sizes $16$, $43$, $124$, $367$, $1096$, and $3283$, respectively. Contrast to
the conclusions in Ref. \cite{Yang}, the SA walk does not reduce the search
time and is even more inefficient than the NTL and NQL walks when $N$ is
large. We argue that what is important for the improvement of the search
efficiency is not taking more strict law to eliminate repeating visits, but
considering the factual structure of the networks.

\begin{figure}
\centerline{\resizebox{8cm}{!}{\includegraphics{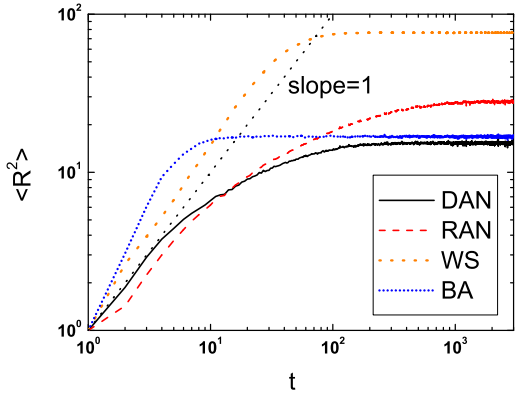}}} \caption{Mean-square
distance $\langle{R^{2}}\rangle$ as a function of time for DANs, RANs, WS and
BA networks with size $N=9844$ and average degree $\langle{k}\rangle=6$ fixed.
The dot line with slope $1$ is plotted for comparison.} \label{fig4}
\end{figure}

\begin{figure}
\centerline{\resizebox{10cm}{!}{\includegraphics{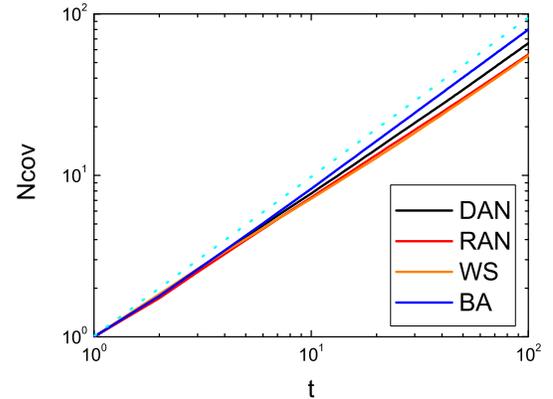}}} \caption{(Color
online) Network coverage $N_{cov}$ after $t$ steps on (solid line from top to
bottom) BA networks, DANs, RANs and WS networks with $N=9844$ and
$\langle{k}\rangle=6$ fixed.} \label{fig5}
\end{figure}

According to the former analysis of MFPT, we know that the high-degree nodes would be
visited early, so as to be avoided by SA walker preferentially. Thus, to some
extent the SA search process can be considered as the question of intentional
attack \cite{AlbertAttack} or preferential site percolation (PSP) on networks.
The intentional attack means the removal of nodes and their incident edges
aiming at those greatest degree nodes, and the PSP threshold $P_{c}$ is a
measure of how robust the network is against this attack. When the network
size is relatively small ($N<10^5$), DANs and RANs are much frailer than SW
and SF networks under the intentional attacks (see Table. \ref{tab:table1}
for $P_{c}$), because of the crucial importance of the high-degree nodes to
the network integrates \cite{Zhou}. When walking on
networks, if the walkers strictly avoid visiting these crucial high-degree
nodes, they can hardly move from their current part of the network to other
parts, as if the network was intentionally attacked and disintegrated into
partes. However, as a search strategy, the SA search processes on the original
network are not totally the same as the intentional attacks, because SA walker
at node $i$ can revisit nodes, including these crucial nodes, although, with
probability $1/k_{i}$ only in condition that all the $k_{i}$ neighbors have
already been visited. The compulsive revisits insure that the targets in
different parts of networks can finally be found. Considering the similarity
between SA search and intentional attack or PSP, we can conclude that the SA
search strategy, compared with other mentioned strategies, will improve the search
efficiency to certain extent for the networks with higher PSP threshold.

Next, we will present our simulation results for two quantities, the
mean-square displacement and the network coverage (the average number of
distinct sites visited). The mean-square displacement
$\langle{R}^{2}(t)\rangle$ of a particle diffusing in a given space, which is a
measure of the distance $R$ covered by a typical RW walker after performing $t$
steps, is one of the most basic quantities in the random walk theory
\cite{Almaas,Gallos}. In most cases, this quantity is described by an
expression of the form $\langle{R}^{2}(t)\rangle\sim{t^{a}}$. The value of the
parameter $a$ classifies the type of diffusion into normal linear diffusion ($a=1$), subdiffusion ($a>1$), or superlinear diffusion ($a<1$). When considering the unrestricted
RW walk along the bonds of networks, the maximum value of $a$ can be $2$ \cite{Gallos}.
To calculate $\langle{R}^{2}(t)\rangle$ we first, at each time step, find the minimal
distance from the current position of the walker to the origin (i.e., the
smallest number of steps needed for the walker to reach the origin) using a
breadth-first search method. Then we allow the walker to move through the
network until $\langle{R}^{2}(t)\rangle$ has saturated. Finally, the results
are average over different initial positions of the walkers and realizations of
the network. We simulate $\langle{R}^{2}(t)\rangle$ for the DAN, RAN, WS and BA
networks with $N=9844$, and report it as a function of time in Fig. \ref{fig4}.
One important feature is the fact that $\langle{R}^{2}(t)\rangle$ equilibrates
after a few steps to a constant displacement value. This is a simple
manifestation of the small diameter of these finite networks. Note also that,
because of the differences of the average path length $\langle{L}\rangle$ of these
networks (see Table. \ref{tab:table1} for $\langle{L}\rangle$), the plateau values are also different. For the DANs and RANs, one can find that the slope of
$\langle{R}^{2}\rangle$ in the early time
is small, especially that of RANs. This can be explained as a result of their
highly clustering effect which induces the walker initially spends many time
exploring the cluster it was created in and thus the distance to the origin
increases slowly. After a few steps the walker escapes its initial territory
and diffuses around the entire network. From the slope of $\langle{R}^{2}\rangle$,
we know that diffusions both on DANs and RANs are sublinear, and diffusions both on BA
and WS networks are superlinear. The network coverage denoted by $N_{cov}$ is
defined as average number of distinct visited nodes of the RW walker. Results
of $N_{cov}$ on the four networks are presented in Fig. \ref{fig5}. The number
of steps performed is nearly two orders of magnitude smaller than the size of
the networks, in order to decrease finite size effects. The clustering effects
of DANs and RANs are reflected by their relatively low coverage compared to BA networks.

In summary, we present comparative studies of the dynamics of random walks on
deterministic Apollonian networks and random Apollonian networks with other
networks. The efficiency of search strategies are
simulated and the clustering and short path effects of networks to the dynamics
are discussed. It is found that, for the DAN with large size, the best
search strategy is no longer the self-avoiding random walk. The importance of
the high-degree nodes to the network integrates are suppose to be one reason.
Thus the search strategy on networks should be assigned closely based on
the topological structure of the underlying networks, including the clustering,
the average path length as well as the essentiality of the high-degree nodes. Since
search is a problem of extreme importance for so many natural and artificial
networks, this finding may be valuable in practicability. Finally, the simulation
results of mean-square displacement $\langle{R}^{2}(t)\rangle$ and network
coverage $N_{cov}$ also show the influences of the structure of networks to
the dynamics of random walks.

\end{document}